\documentstyle[12pt]{article}
\input epsf
\newcommand{\be}{\begin{equation}}
\newcommand{\ee}{\end{equation}}
\newcommand{\bea}{\begin{eqnarray}}
\newcommand{\eea}{\end{eqnarray}}
\newcommand{\lb}{\label}
\begin{document}
\begin{titlepage}
\title{Accelerating Universes With Scaling Dark Matter}
\author{Michel Chevallier$^{1}$ and David Polarski$^{1,2}$\\
\hfill \\
$^1$ Lab. de Math\'ematique et Physique Th\'eorique, UPRES-A 6083\\
  Universit\'e de Tours, Parc de Grandmont, F-37200 Tours (France)\\
\hfill\\
$^2$ D\'epartement d'Astrophysique Relativiste et de Cosmologie,\\
  Observatoire de Paris-Meudon, 92195 Meudon cedex (France)}

\date{\today}

\maketitle

\begin{abstract}
Friedmann-Robertson-Walker universes with a presently large fraction of the energy 
density stored in an $X$-component with $w_X<-1/3$, are considered. 
We find all the critical points of the system for constant equations of state in 
that range. We consider further several background quantities that can distinguish the models 
with different $w_X$ values. Using a simple toy model with a varying 
equation of state, we show that even a large variation of $w_X$ at small redshifts  
is very difficult to observe with $d_L(z)$ measurements up to $z\sim 1$.
Therefore, it will require accurate measurements in the range $1<z<2$ and 
independent accurate knowledge of $\Omega_{m,0}$ (and/or $\Omega_{X,0}$) in order to resolve 
a variable $w_X$ from a constant $w_X$.    
\end{abstract}

\end{titlepage}

\section{Introduction}
Since the release of type Ia Supernovae data independently by two groups, the Supernovae Cosmology Project 
and the High-$z$ Survey Project \cite{Perl,Garn}, indicating that our universe is 
presently accelerating, 
there has been a wealth of activity trying to explain the origin of this acceleration. 
In particular it was found that, assuming flatness, the best-fit universe with a cosmological 
constant $\Lambda$ is given by $(\Omega_{\Lambda,0},~\Omega_{m,0})=(0.72,~0.28)$. 
Though these data have still to be confirmed by more accurate ones, as will be the case in the near 
future, it is clear that, if taken at face value, they imply a radical departure from usual 
textbooks Friedman-Robertson-Walker cosmology. 
Still this possibility is very attractive as it seems to make all pieces of our present 
understanding of structure formation fit into a consistent paradigm \cite{BOPS,SS}.
But a true cosmological constant, interpreted as a vacuum energy, is about $123$ orders of 
magnitude smaller than its ``natural'' value. This is why there has been a lot of interest in models where 
the dominant fraction of the present energy density is rather some 
effective (slowly varying) cosmological constant term. 
The prominent candidate in this respect is some minimally coupled scalar field $\phi$ \cite{RP88,CDS98} 
often called quintessence, slowly 
rolling down its potential such that it can have negative pressure. 
Actually, if this candidate is the right one, then it is in principle possible to reconstruct (the relevant 
part of) its potential $V(\phi)$ \cite{St98,HT99} and the corresponding equation of state $w_{\phi}$ 
\cite{SSSS,AW00} using luminosity distance measurements in function of redshift $z$, 
a procedure that can in principle be extended to so-called 
generalized ``quintessence'' in the framework of scalar-tensor theories of gravity \cite{BEPS00} .
A $\Lambda$-term is equivalent to a perfect isotropic fluid with a constant equation of state 
$p=w \rho,~w=-1$.
We will be interested in this work in some dominant $X$-component, which can be described as 
a perfect fluid with an equation of state specified by $-1<w_X\equiv p_X/\rho_X<-1/3$, 
thus enabling that component to induce accelerated expansion.  
We compare these models, in particular those in the range of values in best agreement with 
observations, $-1<w_X<-0.6$, with $\Lambda$ dominated universes. We don't consider here 
constant equations of state with $w_X<-1$, though this even more mysterious possibility is not 
excluded by observations \cite{Cald} (however we will consider such cases for variable $w_X$). 
It is possible to have a constant equation of state, even in the framework of quintessence though this 
implies severe constraints on the potential $V(\phi)$ \cite{SS,DPD00}. Anyway, results obtained here 
apply to these 
cases if the $X$-component is some minimally coupled scalar field. Though a constant $w_X$ is a 
simplified assumption, these models give insight into the observational properties of more 
sophisticated models. 
Finally, we also consider a toy model with a variable equation of state where the variation of $w(z)$ 
is significant in the range $0\le z\le 2$. 
An arbitrary $V(\phi)$ is expected to give a variable equation of state. Our aim is, particularly 
in view of present experimental and theoretical activity, to see whether 
this equation of state which has non negligible variation in the range $0<z<2$ can be unambiguously 
separated from models with a constant equation of state. We show that the separation 
of these models using luminosity distance measurements will require accurate measurements in the 
range $1\le z\le 2$, to be probed in the future by the SNAP satellite, a conclusion that can be evidently 
extended to all viable models with variable equation of state.  

\section{FRW cosmologies}
Our starting point are the well-known Friedmann equations for a homogeneous and isotropic 
universe,
\bea
\biggl(\frac{\dot a}{a}\biggr)^2 + \frac{k}{a^2} &=& \sum_{i} \frac{8\pi G}{3} ~\rho_i\lb{FR1}\\
\frac{\ddot a}{a} &=& -\frac{4\pi G}{3}(\rho_i + 3 p_i)\lb{FR2}
\eea
Here the different matter components labelled $i$, are all isotropic perfect fluids. 
We note from (\ref{FR2}) that a component $i$ can induce accelerated expansion provided 
$\rho_i + 3 p_i<0$.
We are interested first in equations of state of the form $\rho_i=w_i~p_i$, with constant $w_i$.
It is convenient to write these equations using the dimensionless quantities 
$x\equiv \frac{a}{a_0}, ~T\equiv|H_o|~t$ and $\Omega_{i,0}\equiv \frac{\rho_{i,0}}{\rho_{cr,0}}$ 
where $H=\frac{\dot a}{a},~\rho_{cr,0}\equiv \frac{3H_0^2}{8\pi G}$ and $f_0$ stands for any 
quantity $f$ evaluated today (at time $t_0$), finally $\Omega_{k,0}\equiv -\frac{k}{a_0^2H_0^2}$.  
Note that this parameterization fails when $H_0$ vanishes.
Equations (\ref{FR1},\ref{FR2}) are then rewritten as 
\bea
\frac{1}{2} \biggl( \frac{dx}{dT}\biggr)^{2}\equiv \frac{1}{2} {\dot x}^2 
&=&\frac{1}{2} \sum_{i} \Omega_{i,0}~x^{-(1+3w_{i})} +\frac{1}{2} \Omega_{k,0}\lb{T1} \\
\frac{d^{2}x}{dT^{2}}\equiv {\ddot x} 
&=& - \frac{1}{2} \sum_{i} \Omega_{i,0}~(1+3w_{i})~x^{-(2+3w_{i})}\lb{T2}
\eea
We see in particular that $x_0=1,~{\dot x}_0 = 1~{\rm or}~-1$. Further $\ddot{x}_0=-q_0$, 
where $q_0$ is the deceleration parameter.
Therefore, we have $\sum_{i} \Omega_{i,0}+\Omega_{k,0}=1$ (actually this equality can be 
extended to any given time). We could in principle include the curvature term 
$\frac{1}{2} \Omega_{k,0}$ into the sum in the r.h.s. of (\ref{T1}), it would correspond to 
$w_k=-1/3$, but we prefer to keep it separate for better physical insight, 
also this curvature term does not appear in (\ref{T2}).
It is suggestive to rewrite (\ref{T1},\ref{T2}) further as 
\bea
\frac{1}{2} \dot{x}^{2} &=& E-V(x)\\
\ddot{x}&=& -\frac{dV}{dx}~,
\eea
with the following identifications 
\bea
V(x)&=&-\frac{1}{2} \sum_{i} \Omega_{i,0}~x^{-(1+3w_{i})}\\
E&=&\frac{1}{2} \Omega_{k,0}~.
\eea
In the Hamiltonian formulation of our problem, we have 
\be
H = \frac{p_x^2}{2} + V(x)~,
\ee
where $p_x$ is the momentum conjugate to $x$, with the usual equations of motion 
\be
{\dot x} = \frac{\partial H}{\partial p_x}~~~~~~~~~~~~~~~~~~~~~
{\dot p_x} = -\frac{\partial H}{\partial x}~.  
\ee
Our problem is therefore similar to that of a particle in one dimension with potential 
energy $V(x)$. This gives at once insight into the possible motions, like in problems considered 
in classical mechanics.  
In the following we consider a universe filled with dustlike matter ($p=0)$ and some unknown component 
labelled $X$ with negative pressure, $-1\le w_X<-1/3$. Hence we have $\rho_X + 3 p_X<0$.
A pure positive cosmological constant, i.e. a $\Lambda$-term, corresponds to $w_X=-1$. 

Present observations indicate $\Omega_{X,0}\sim 0.7$ and though this is large compared to dustlike 
matter, $\rho_{X,0}$ is of course negligible compared to ${M_{pl}}^4$, the natural order of magnitude 
of the vacuum energy. This is why there are attempts to explain the $X$-component in terms of 
some scalar field $\phi$, a mechanism quite similar to inflation. It is possible in that case 
too to have scaling solutions with $\rho_{\phi}\propto x^m,~m=-3(1+w_{\phi})$=constant. However 
this requires a very particular potential $V(\phi)$ for which, 
\be
V(\phi) = \frac{1-w}{1+w}~\frac{\dot \phi^2}{2}~.
\ee

Clearly, when $w_X<-1/3$, the energy density 
$\rho_X$ decreases even slowlier than the curvature term $\propto x^{-2}$ and, as is the case for a 
cosmological constant, interesting new features appear in the dynamics of our system. 
We drop in the following the relativistic component as it is negligible today and basically of no 
relevance for the 
issues considered in this work.
Hence 
\bea
{\dot x}^2 &=& \Omega_{m,0} ~x^{-1}+
\Omega_{X,0} ~x^{-(1+3w_X)} + \Omega_{k,0} \\
\ddot{x} &=& -\frac{1}{2} \frac{\Omega_{m,0}}{x^{2}}-\frac{1}{2}\Omega_{X,0} ~(1+3w_{X}) 
~x^{-2-3w_{X}}\lb{q}~,
\eea
and we have readily the identity
\be
V(x) = -\frac{1}{2}\Omega_{m,0} ~x^{-1} - \frac{1}{2}\Omega_{X,0} ~x^{-(1+3w_X)}
\ee
From (\ref{q}), our universe is presently accelerating provided 
\be
w_{X}<-\frac{1}{3} \left( 1 + \frac{\Omega_{m,0}}{\Omega_{X,0}} \right)~,
\ee
and in particular for a flat universe
\be
w_{X}<-\frac{1}{3} \Omega_{X,0}^{-1}~.
\ee
For instance, for $\Omega_{m,0}=0.3,~\Omega_{X,0}=0.7$, $w_X<-0.47$ is required. Therefore, present 
experimental estimates based on baryons in clusters on one hand giving $\Omega_{m,0}\sim 0.3$, 
and on the location of the first 
acoustic peak in the Cosmological Microwave Background detected by the balloon experiments Boomerang and 
Maxima \cite{Bo,Ma} suggesting a nearly flat universe, on the other hand, imply that our universe 
would be presently accelerating for a wide range of constant values, roughly $w_X<-0.5$. 
Accelerated expansion starts at $x_{a}$ given by 
\be
x_{a}^{-3 |w_X|}= (-1+ 3|w_X|) ~\frac{\Omega_{X,0}}{\Omega_{m,0}}~,\lb{xa}
\ee
which corresponds to redshifts 
\be
z_a = (-1+ 3|w_X|)^{\frac{1}{3|w_X|}} ~\left( \frac{\Omega_{X,0}}{\Omega_{m,0}} \right)^{\frac{1}{3|w_X|}} 
- 1~.\lb{za} 
\ee
For $-1<w\to -1/3,~z_a$ is shifted towards smaller redshifts, for 
$(\Omega_{m,0},~\Omega_{X,0})=(0.3,~0.7)$, we have $z_a=0.671$ for a constant $\Lambda$-term, and 
$z_a=0.414$ when $w=-0.6$. The fact that $z_a$ is so close to zero, is the cosmic coincidence problem.
 
Let us look now for the critical points of our system, points for which ${\dot x_{cr}} = {\dot p_{cr}} =0.$
While, by definition, points $x_a$ introduced in (\ref{xa}) have no acceleration, the absence of 
expansion (or contraction) requires in addition 
\be
E - V(x_a) = 0~.\lb{cr}
\ee  
Any particular point $x_a$ satisfying also (\ref{cr}) will yield a critical point $x_{cr}$.
Hence, the existence of a critical point translates into the following equation for the parameters 
$\Omega_{X,0},~\Omega_{m,0}$
\be
y^{3|w_X|} + \alpha y + \beta = 0~,~~~~~~~~~~~~~~y\ge 0 \lb{y}
\ee   
where 
\bea
y &\equiv& \biggl( \frac{\Omega_{X,0}}{\Omega_{m,0}} \biggr)^{\frac{1}{3|w_X|}}\\
\beta &\equiv& 1 - \Omega_{m,0}^{-1}\\
\alpha &\equiv& - 3 |w_X| ~(-1 + 3|w_X|~)^{\frac{1-3|w_X|}{3|w_X|}}~.
\eea
Note that this equation cannot be used in the particular case $\Omega_{m,0}=0$.
For a pure $\Lambda$-term, eq.(\ref{y}) is cubic.
We note first that $\beta$ is bounded from above, $-\infty<\beta\le 1$ (the equality $\beta=1$ is 
unphysical, it corresponds to $\Omega_{m,0}=\infty$). The polynom in (\ref{y}) has a minimum 
in $y_{min}$ 
\be
y_{min} = (-1+ 3|w_X|)^{-\frac{1}{3|w_X|}}
\ee
and it is easily verified that, for any given value of $\beta$, or equivalently of $\Omega_{m,0}$, 
(\ref{y}) has always at least one solution. We also have the nice equality
\be
x_a=\frac{y_{min}}{y}~.
\ee
The following cases can be distinguished:
\vskip 10pt
\par\noindent
a) $\beta < 0~~(0 < \Omega_{m,0} < 1)$: one critical point is found with $x_{cr}<1$.
\vskip 10pt
\par\noindent
b) $\beta = 0~~(\Omega_{m,0} = 1)$: there are two critical points. The first one corresponds to 
$\Omega_{X,0}=0$ and the critical point is just $x_{cr}=\infty$. 
The second one is for $y$ a solution of $y^{3|w_X|} + \alpha y=0$ and $x_{cr}<1$.    
\vskip 10pt
\par\noindent
c) $0 < \beta < 1~~(1 < \Omega_{m,0} < \infty$): two critical points can be found with $x_{cr}$ 
respectively larger and smaller than $1$.  
\vskip 10pt
\par\noindent
d) $\beta = 1~~(\Omega_{m,0} = \infty)$: this point is clearly unphysical, however we can consider 
the limit $\beta\to  1$. In that case, $x_{cr}\to 1$ (for a cosmological constant this solution tends 
to the static Einstein universe). 
\vskip 10pt
\par
\begin{figure}
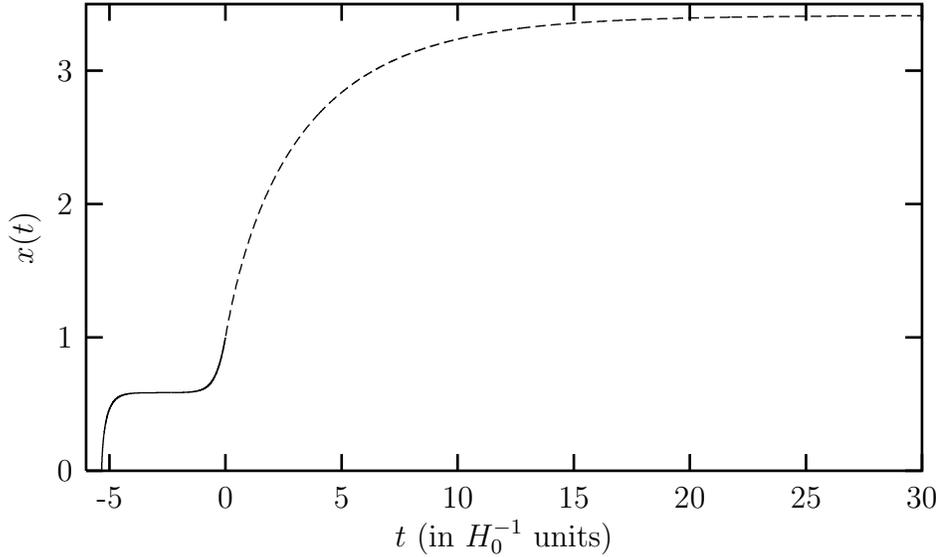

\input figure.tex
\caption[]{For $w_X=-2/3$ and $\Omega_{m,0}=2$, two different universes having 
a quasi-static (loitering) stage are shown. The two universes have different values $\Omega_{X,0}$. 
The present time is chosen here at $t=0$ and we show for each universe the relevant part, 
either $t<0$ or $t>0$, containing a quasi-static stage. 
The quasi-static stages take place in the vicinity 
of the critical points corresponding to $\Omega_{X,0}=5.8284...$ and $\Omega_{X,0}= 0.17157...$.
For an expanding universe with $\Omega_{m,0}=2,~\Omega_{X,0}=5.8284...$ (plain line), loitering 
is in the past and occurs at $z\approx 0.70$ while it is in the future at $x\approx 3.41$ for an 
expanding universe with $\Omega_{m,0}=2,~\Omega_{X,0}= 0.17157...$ (dashed line). 
We could also have a bounce in the past for slightly
higher values of $\Omega_{X,0}$; the other loitering stage can be followed either by expansion 
(this is the case here, but the expansion stage is not shown) or by a recollapse for slightly 
lower $\Omega_{X,0}$ values.}
\label{xcr}
\end{figure}

\noindent
An intriguing possibility when there is a constant $\Lambda$-term is a quasi-static stage first 
discussed by Lema\^itre \cite{L27} (called ``loitering'' stage in \cite{SFS92}).
This stage is followed either by recontraction or eternal expansion. 
Actually such quasi-static stages just correspond to configurations in the vicinity of a critical point.   
Hence such a stage is equally possible when $-1<w_X<-1/3$ and it can take place at redshifts 
\be
z\approx (-1+ 3|w_X|)^{\frac{1}{3|w_X|}}~y - 1 = \frac{y}{y_{min}}-1~,\lb{zl}
\ee
where $y$ is a solution of (\ref{y}). For $\Omega_{m,0}<1$, a quasistatic stage 
in an expanding universe is only possible in the past in contrast to expanding universes with 
$\Omega_{m,0}>1$, where a loitering stage is possible either in the past or in the future, 
and two such universes with $\Omega_{m,0}=2$ are shown in Fig.\ref{xcr}.
Let us finally mention that spatially flat universes, which are particular subcases of $\beta\le 0$,  
can have critical points either in the 
infinite past for $(\Omega_{X,0},~\Omega_{m,0})=(1,~0)$ or in the infinite future for 
$(\Omega_{X,0},~\Omega_{m,0})=(0,~1)$. 
In the latter case it will take an infinite time for the expansion to relax at $x_{cr}=\infty$ while in 
the other case it would take a contracting universe an infinite time to relax at $x_{cr}=0$.
Hence, in a flat expanding universe, and generally in any expanding universe with 
$\Omega_{m,0}<1$, the expansion is eternal for $-1\le w_X=~{\rm constant}<-1/3$.

\section{Some observational constraints}
It is interesting to see how various models fare when compared with observations. 
We consider in the following several quantities probing the background dynamics that can discriminate, 
and possibly rule out, the different types of universes from each other.  
\vskip 10pt
\par
\begin{figure}
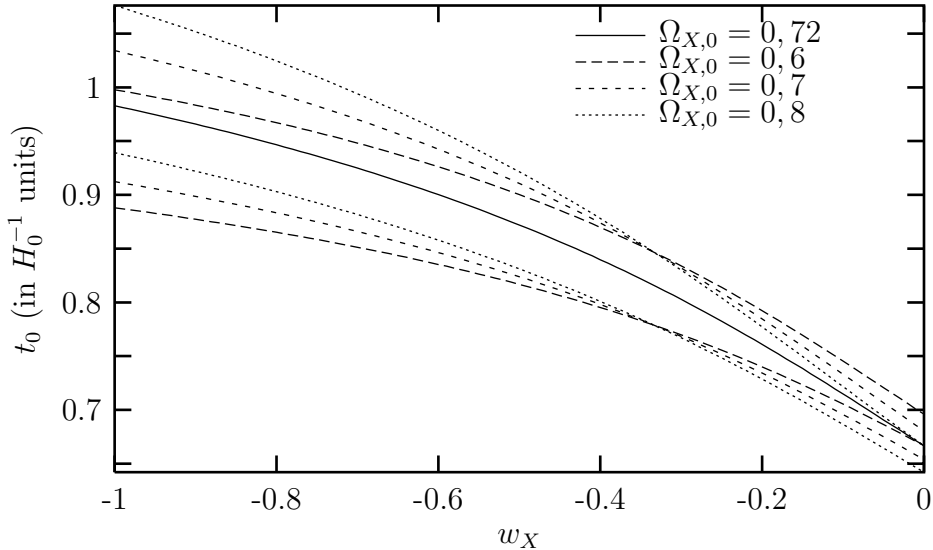

\input figure2.tex
\caption[]{The age of the universe is shown for various cases in function of $w_X=\frac{\rho_X}{p_X}$. 
The central curve corresponds to a flat universe with ($\Omega_{X,0},~\Omega_{m,0})=(0.72,~0.28)$. 
For each $\Omega_{X,0}$, the upper, resp. lower, curve corresponds to $\Omega_{m,0}=0.2$, resp., $0.4$.
Note that all universes with same $\Omega_{m,0}$  have the same age when the $X$-component scales like 
the curvature term.}
\label{agew}
\end{figure}

\noindent
{\it Age of the Universe}: This was actually the main incentive for reconsidering the introduction 
of a cosmological constant. Indeed, it is possible for an old universe to accomodate with a high value 
of $H_0$. 
The universe with $(\Omega_{\Lambda,0},~\Omega_{m,0})=(0.72,~0.28)$ and 
$0.6\le h\equiv H_0/100~{\rm km/s/Mpc}\le 0.7$ has an age 
$13.7 \le t_0\le 16$ Gyrs, an increase of nearly 50\% compared to a flat matter-dominated universe!
When we change the value of $w_X$, keeping the other cosmological parameters fixed, the age will 
decrease with increasing values of $w_X$. 
%
%
As can be seen from Fig.\ref{agew}, varying for example 
$\Omega_{X,0}$ and $w_X$, we have a degeneracy for the age of the universe for some values of 
$\Omega_{m,0}$ compatible with observations. Therefore the constraint of a minimum age $t_0$ will 
select a region in the free parameter space (for example the $\Omega_{X,0}-w_X$ plane if we keep 
$\Omega_{m,0}$ fixed). We see also the 
effect of a varying $w_X$ when $\Omega_{m,0},~\Omega_{X,0}$ are kept constant. Note the particular value 
$w_X=-1/3$ for which all universes with same $\Omega_{m,0}$ give the same age as can be easily 
checked with the analytic expression for $t_0$.
\be
t_0 = \int_0^1 \frac{dx}{x~H(x)}~.
\ee
Of course, we are interested in values of $w_X$ sensibly smaller than $-1/3$.  
\vskip 10pt
\par
\begin{figure}
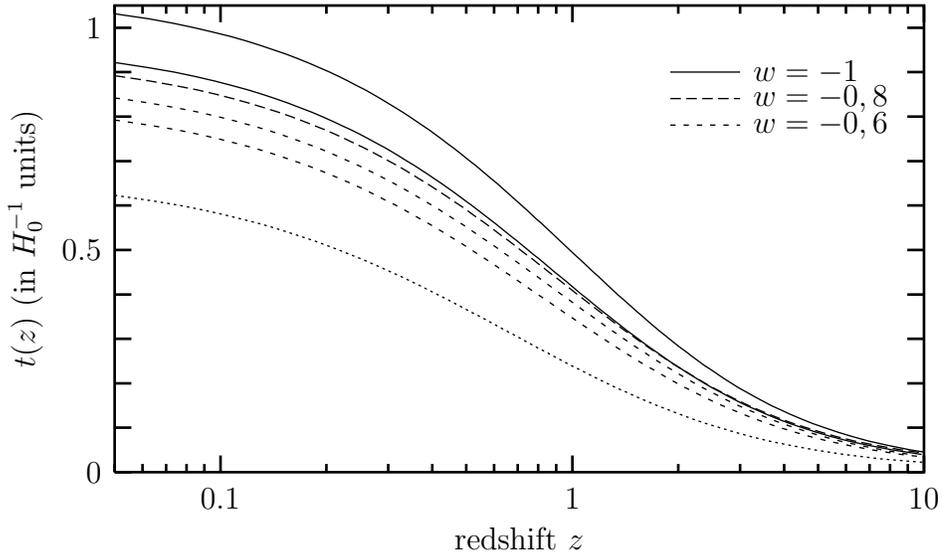

\input figure5.tex
\caption[]{The age of the universe at various redshifts $z$ is shown for several universes. The upper 
curve has $\Omega_{m,0}=0.2$, the lowest dashed curve (with $w_X=-0.6$) has $\Omega_{m,0}=0.4$. 
The three curves in between all have $\Omega_{m,0}=0.3$. All these universes pass the constraint 
set by high-redshift objects 
mentioned in the text for $0.6\le h\le 0.7$. However the lowest dashed curve, is already on the 
border for h=0.7. We show for comparison $t(z)$ for the Einstein-de Sitter 
universe (dotted line).}
\label{tz}
\end{figure}

\noindent 
{\it Age in function of redshift}: this constraint is more detailed than the previous one as it puts 
an age limit at each redshift. 
This quantity has the following integral expression 
\be
t(z) = \int_0^{(1+z)^{-1}} \frac{dx}{x~H(x)}~.
\ee
An ``old'' object observed at a certain redshift selects all models 
with at least that age at that given redshift. In that respect, several age constraints have recently 
appeared in the literature \cite{D96}. For example, the age of the radio galaxy 53W091 observed at a redshift 
$z=1.55$ puts a lower bound of $3.5$Gyrs at that redshift. The quasar observed at $z=3.62$ sets 
a lower bound of $1.3$Gyrs. We see from Fig.\ref{tz} that all universes displayed, besides the flat 
matter-dominated universe, can accomodate these 
constraints for a value $0.6\le h\le 0.7$. However a flat purely matter dominated universe 
cannot, unless the Hubble constant $H_0$ is lower than the present estimate $h=0.65\pm 0.05$.  
If we vary only $w_X$, larger values give lower $t(z)$, hence accurate age dating of far objects 
can possibly set a maximal value for $w_X$.  
\vskip 10pt
\par
\begin{figure}
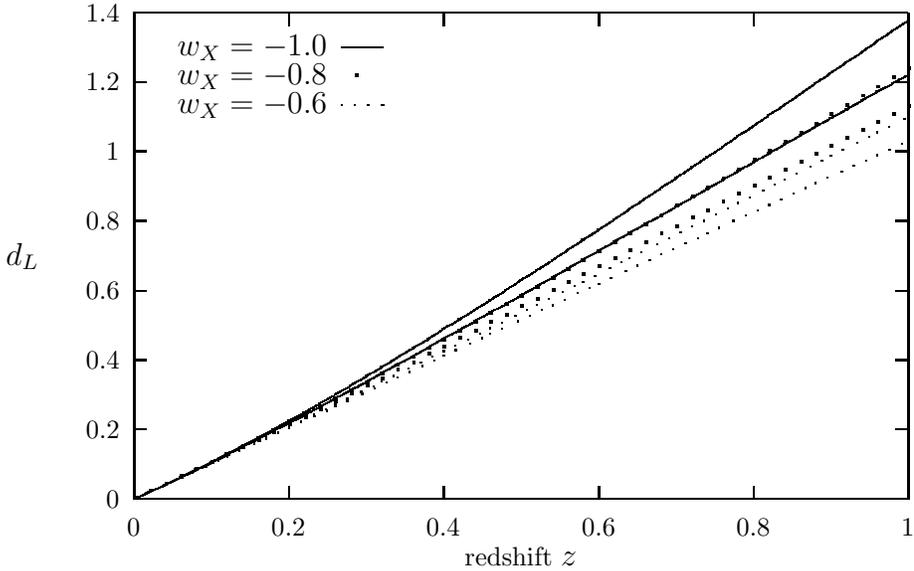

\input figure3a.tex
\caption[]{The luminosity distance (in $H_0^{-1}$ units) is plotted against redshift for several 
flat universes. For each constant $w_X$ value, the upper, resp. lower, curve corresponds to 
$\Omega_{m,0}=0.2$, resp. $0.3$. We see that some combinations conspire to give practically 
undistinghishable curves up to redshifts $z\sim 1$.}
\label{dlz}
\end{figure}

\noindent
{\it The luminosity distance $d_L(z)$}: it is essentially this quantity (through the measurement of 
the Hubble diagram) 
which was measured for type Ia Supernovae at redshifts up to $z\sim 1$ by two teams, 
the Supernovae Cosmology Project and the High-$z$ Survey Project.
These two teams have found evidence that our universe is presently accelerating, assuming flatness and 
$w_X=w_{\Lambda}=-1$, a best-fit universe is obtained for 
$(\Omega_{m,0},~\Omega_{\Lambda,0})=(0.28,~0.72)$.
We have the integral expression 
\be
d_L(z) = (1+z)~H_0^{-1}~|\Omega_k|^{-\frac{1}{2}}~{\cal S}\left( H_0~|\Omega_k|^{\frac{1}{2}}~
\int_{(1+z)^{-1}}^1 \frac{dx}{x^2~H(x)}\right)~,
\ee
with ${\cal S}(u)=\sin u$ for a closed universe, ${\cal S}(u)=\sinh u$ for an open universe 
and ${\cal S}$ is the identity for a flat universe. 
In the near future, several groups will gather data for samples of around a hundred 
supernovae each and the SNAP (Supernova Acceleration Probe) satellite is planned to make 
measurements on thousands supernovae with an accuracy at the percentage level 
up to redshifts $z=1.7$ \cite{SNAP}.
In Fig.\ref{dlz} 
we see 
the effect of varying the equation of state and/or the balance between the parameters 
$\Omega_{m,0},~\Omega_{X,0}$ for flat universes. 
Several flat models are considered and some models are easily resolved but for small redshifts 
they are still very close.
In particular, we see that different combinations of $\Omega_{m,0}$ and $w_X$, can conspire to give 
practically the same $d_L(z)$ up to redshifts $z\sim 1$.
Therefore high accuracy measurements with uncertainties at the percentage level are needed in 
order to cleanly distinguish the models and the need to go to redshifts sensibly 
higher than $1$ is evident. But equally crucial, as canalso be seen, one needs independent 
knowledge of $\Omega_{m,0}$ and $\Omega_{X,0}$ (only one of them if we assume flatness) with 
uncertainties much smaller than $0.1$.
%
%
To see the effect of a variation of the equation of state, we compare several models sharing 
the same value $w_X(z=0)$.
As we have said above, if we want to explain the $X$-component using a scalar field, 
it is natural to expect a time varying $w_X$. 
Clearly, if we are able to pinpoint the equation of state, crucial information will be gained about the 
underlying particle physics model. 

Let us consider a toy model in which a 
phenomenological $w_X(z)$ is assumed and we want to consider in how far such a model can be 
distinguished from a model with constant $w_X$. We take the following simple model:
\be
w_X=-1 + \alpha + \beta (1-x)~.\lb{w}
\ee 

\begin{figure}
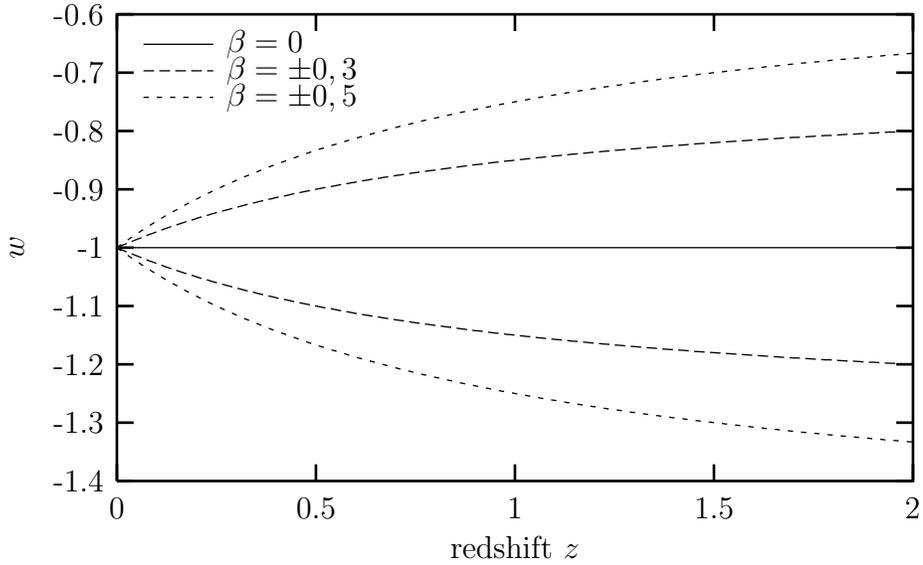

\input figure7.tex
\caption[]{The value $w_X(z)$ is shown for our toy model. Note that we display here the cases with 
$\alpha=0$ which correspond to the present value $w_X(0)=-1.$}
\label{wz}
\end{figure}

The two constants $\alpha$ and $\beta$ are chosen arbitrarily. The parameter $\alpha$ measures the 
departure today from an equation of state corresponding to a cosmological constant, while the 
parameter $\beta$ measures the variation in time of $w_X$. 
Adopting a purely phenomenological approach, we will consider also cases where $w_X$ becomes 
smaller than $-1$ which can not be accounted for by a minimally coupled scalar field.  
The energy density $\rho_X$ evolves in time according to  
\be
\rho_X = \rho_{X,0}~x^{-3(\alpha + \beta)}~e^{-3\beta(1-x)}~.
\ee
\begin{figure}
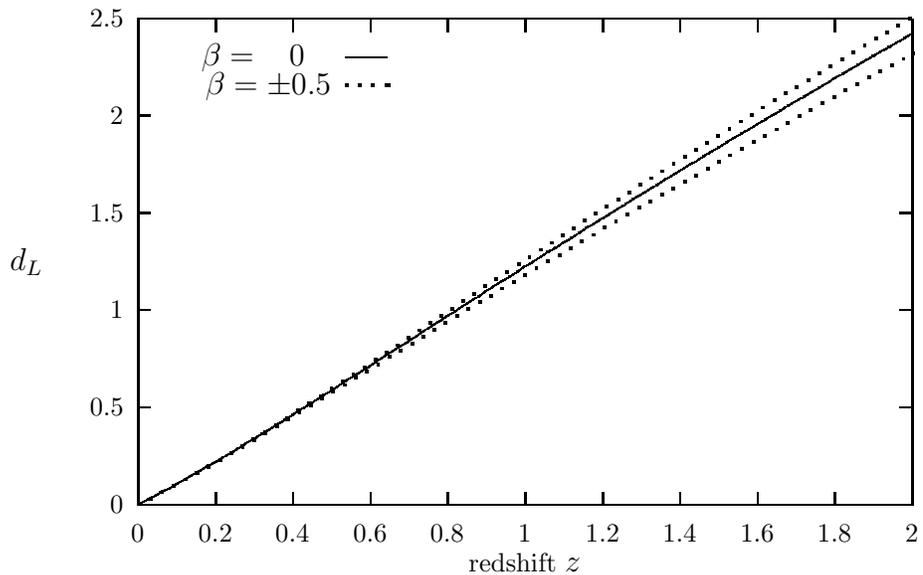

\input figure6a.tex
\caption[]{The luminosity distance (in $H_0^{-1}$ units) is shown for some of the variable equations of 
state displayed on the previous Figure and compared to a pure cosmological constant ($\beta=0$). 
Flatness is assumed with $\Omega_{m,0}=0.3$. Obviously, accurate measurements at redshifts $1<z<2$, 
and accurate knowledge of $\Omega_{m,0}$ are needed in order to separate the three cases.}
\label{dlz1}
\end{figure}
From Fig.\ref{dlz1}, it is evident that even a strongly varying $w_X$ does not yield a significant 
difference compared to the case with $w_X$=constant. However, at redshifts $z\sim (1-2)$, a measurement 
with an accuracy at the percentage level might be able to distinguish our varying equation of state 
with $\beta=0.5$ from a constant equation of state (and same value today). However, for 
$\beta\sim 0.1$, clean separation will become a technological challenge.  

We see that a variation of the cosmological parameters leads to a better separation of the 
models than a variation in the equation of state, therefore we need independent accurate 
knowledge of $\Omega_{m,0}$ (assuming a flat universe) with an uncertainty 
$\Delta \Omega_{m,0}$ significantly smaller than $0.1$ in order to resolve the various curves.
A detailed discussion of the feasibility of this resolution, of the use of fitting 
functions, and of the reconstruction of the equation of state, is beyond the scope of this work.   



\end{document}